# Revealing nanomechanical domains and their transient behavior in mixed-halide perovskite films

Ioanna Mela[a#], Chetan Poudel[a#], Miguel Anaya[b]*, Géraud Delport[b], Kyle Frohna[b], Stuart Macpherson[b], Tiarnan A. S. Doherty[b], Samuel D. Stranks[a,b]*, Clemens F. Kaminski[a]*

Dr I. Mela, Dr. C. Poudel, Dr. S. D. Stranks, Prof. C. F. Kaminski
Department of Chemical Engineering & Biotechnology, University of Cambridge, Philippa Fawcett Drive, Cambridge CB3 0AS, UK.
cfk23@cam.ac.uk

Dr. M. Anaya, Dr. G. Delport, Kyle Frohna, Stuart Macpherson, Tiarnan A. S. Doherty, Dr. S. D. Stranks
Cavendish Laboratory, University of Cambridge, JJ Thomson Avenue, Cambridge CB3 0HE, UK.
ma811@cam.ac.uk
sds65@cam.ac.uk

# These authors contributed equally.

**Abstract** Halide perovskites are a versatile class of semiconductors employed for high performance emerging optoelectronic devices, including flexoelectric systems, yet the influence of their ionic nature on their mechanical behaviour is still to be understood. Here, a combination of atomic-force, optical and compositional X-ray microscopy techniques is employed to shed light on the mechanical properties of halide perovskite films at the nanoscale. We reveal mechanical domains within and between morphological grains, enclosed by mechanical boundaries of higher Young's Modulus than the bulk parent material. These mechanical boundaries are associated with the presence of bromide-rich clusters as visualized by nano-X-ray fluorescence mapping. Stiffer regions are specifically selectively modified upon light soaking the sample, resulting in an overall homogenization of the mechanical properties towards the bulk Young's Modulus. We attribute this behaviour to light-induced ion migration processes that homogenize the local chemical distribution, which is accompanied by photobrightening of the photoluminescence within the same region. Our work highlights critical links between mechanical, chemical and optoelectronic characteristics in this family of perovskites, and demonstrates the potential of combinational imaging studies to understand and design halide perovskite films for emerging applications such as photoflexoelectricity.

Since 2009, halide perovskites have emerged as promising materials for efficient optoelectronic devices such as photovoltaics (PV) and light-emitting diodes (LEDs)[1–3], with performance in some configurations already exceeding comparable existing commercial technologies[4,5]. Halide perovskite semiconductors show unique properties such as remarkable defect tolerance[6], efficient light absorption[7,8] and long charge-carrier diffusion lengths[9–11], which put them at the forefront of emerging thin-film technologies. One advantage they have over conventional semiconductors is that they can be fabricated through facile and cost-effective preparation techniques (e.g. solution processing), with the precursor components self-assembling into relatively large entities (domains of ca 0.1 to several micrometers in size [12], referred to as grains). Moreover, the organic-inorganic hybrid nature of these materials enables a wide variety of possibilities in terms of compositional management[13,14] to tailor their properties to a given application. Variability associated with current preparation techniques combined with this compositional tuning give rise to complex heterogeneity across multiple length scales[15–17], including variations in morphological, structural and photophysical properties.

In particular, state-of-the-art halide perovskite devices are based on multi-cation (methylammonium, MA; formamidinium, FA; Cesium, Cs) mixed-halide (Bromide, Br; Iodide, I) compositions[18], such as the 'triple cation' composition $Cs_{0.05}MA_{0.17}FA_{0.78}Pb(I_{0.83}Br_{0.17})_3$. The alloyed nature of these films tends to generate local variations of the chemical composition at the nanoscale, as observed by electron and X-ray microscopy techniques[19–22]. These compositional heterogeneities correlate directly with crystallographically distinct entities, generating nanoscopic interfaces where discrete deep trap clusters are prominent[16]. Control over the resulting compositional variations represents a promising strategy to tailor transport of excited species[3] or photodoping[23] and make use of the full potential of these alloyed halide perovskites. Photoinduced chemical reorganization of the halides adds another layer of complexity that can either be problematic for operation[24], or desirable, for example to form composition gradients or to passivate defects[25,26]. Additionally, recent reports suggest that strain gradients in these materials play a crucial role in the final performance of devices[27,28], which, in turn, may show a strong flexoelectric response substantially boosted by photoexcitation[29]. For these reasons, understanding the mechanical properties of these semiconductors and their link to compositional variations, light exposure, and thus photophysical performance is generating widespread interest in the research community.

Halide perovskites are substantially softer materials[30,31] (Young's Modulus (YM) below 20 GPa) than purely inorganic semiconductors such as Si, GaAs[32], CIGS[33] or other oxide perovskites[33] (YM on the order of hundreds of GPa). This softness suggests that applying small mechanical stresses might lead to a significant motion of ions in the crystal, and vice versa, through, for instance, photoinduced chemical changes. Indeed, indentation measurements have demonstrated that the YM of single halide perovskites strongly depends on the strength of the Pb-halide bond[33–35]. This observation may translate into nanoscale variations of the mechanical properties in more complex mixed-halide perovskites, depending on their local halide distribution[26]. However, access to nanoscopic mechanical information is elusive for traditional indentation methods, and thus there remains a limited understanding of local mechanical properties in state-of-the-art mixed-halide perovskites and their behaviour under light exposure. Such an understanding is however critical for further device improvements and for the realization of new device paradigms, including straintronics[36,37] or flexoelectric devices[38].

Here, we use quantitative nanomechanical mapping (QNM) atomic force microscopy (AFM) to simultaneously map the morphology and quantify the local YM of triple-cation, mixed-halide $Cs_{0.05}MA_{0.17}FA_{0.78}Pb(I_{0.83}Br_{0.17})_3$ perovskite films. These measurements reveal unexpected variations in the YM, as large as one order of magnitude over just a few tens of nanometres. We observe distinct high YM inter- and intra-grain features, which are identified as bromide-rich regions by nano X-ray fluorescence (nXRF) chemical mapping, and define mechanical domains. Correlative fluorescence lifetime imaging microscopy (FLIM, see Methods for technical details of the experiment) shows an overall increase of the charge carrier lifetime upon light soaking, which is accompanied by a decrease of the YM of of the stiffer entities towards the bulk YM values. This observation suggests that the ionic character confers these materials with transient mechanical behaviour at the nanoscale under *operando* conditions, impacting their photophysical performance. Control over these properties will be critical to stabilizing halide perovskites for their integration in emerging optoelectronic devices, especially those involving both light and mechanical phenomena.[39]

$Cs_{0.05}MA_{0.17}FA_{0.78}Pb(I_{0.83}Br_{0.17})_3$ perovskite thin films were fabricated on glass coverslips following standard solution processing methods (see Experimental Section for details and Figures S1-S3 for morphological, structural and steady-state photoluminescence, PL, characterization). We employ a modular setup to extract mechanical (QNM-AFM mode) and photophysical (FLIM mode) properties of the same region of interest in our films. We follow a sequential methodology between these modes is followed to always map the same perovskite-air interface to ensure appropriate comparisons (see Experimental Section and Figure S4 for an explanation on the methodology). QNM-AFM allows to simultaneously acquire high-resolution (10x10 $nm^2$) topography (**Figure 1**a) and YM (Figure 1b) maps of the perovskite thin film (see Experimental Section and Figure S5). An advantage of this approach compared to traditional indentation is that QNM-AFM does not plastically deform the sample, and therefore does not induce changes in the perovskite material during measurements[40–42] (Figure S6). In addition, we use the FLIM mode of our setup to obtain diffraction-limited PL images across the same area of the sample, providing valuable information about the local dynamics of charge carriers in the perovskite films even though its spatial resolution is lower than that of AFM. A typical FLIM scan is shown in Figure S7 and reveals a mean PL lifetime of 154.4 ± 43.7 ns, in agreement with previous observations for similar excitation fluences [43,44] and also an indication of the high quality of the fabricated films (see Experimental Section).

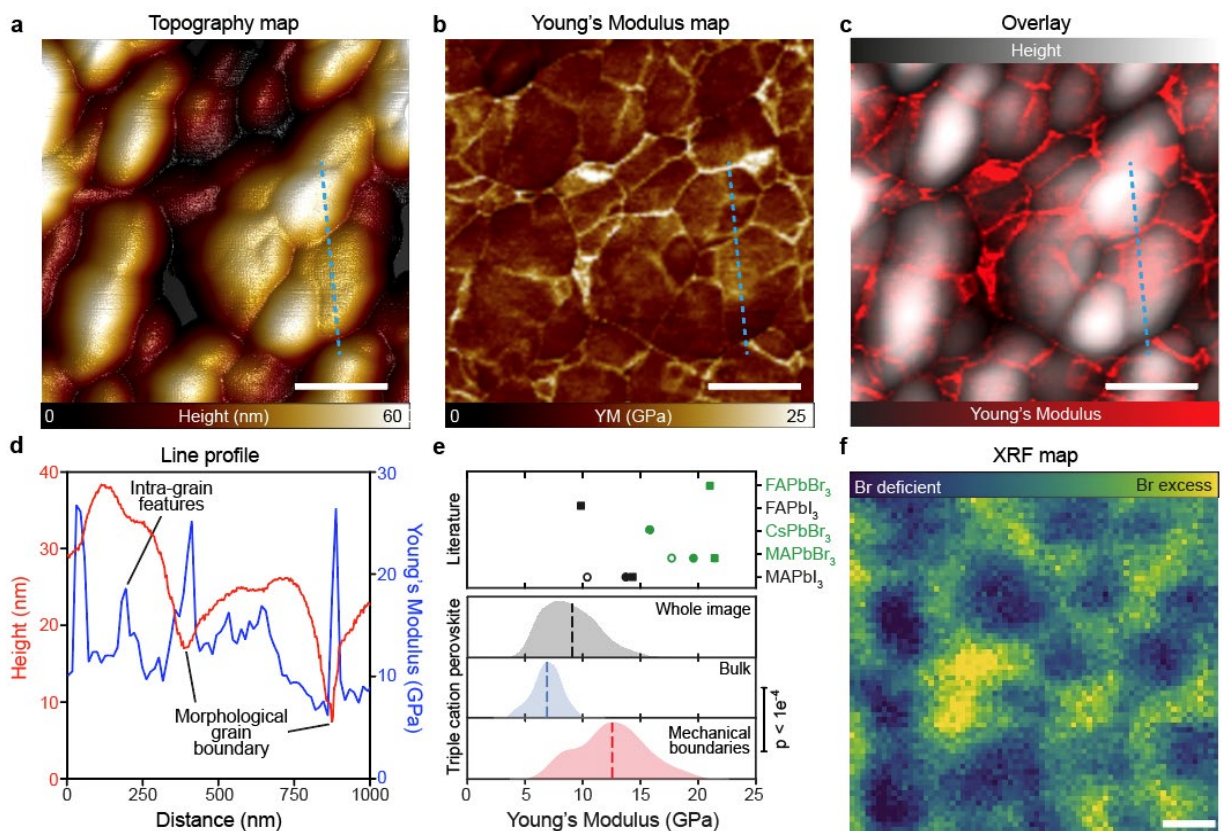

**Figure 1.** QNM-AFM mapping of $Cs_{0.05}MA_{0.17}FA_{0.78}Pb(I_{0.83}Br_{0.17})_3$ perovskite thin films. **a)** Topography maps of the perovskite film show distinct morphological grains. **b)** YM map of the same sample area. **c)** Overlay of the Topography (a) and YM (b) maps, revealing the existence of subdomains of different stiffness within individual morphological grains. **d)** Line profile across two morphological grains (highlighted with blue dotted line in figure 1a-c) shows the height profile (in red) and the Young's modulus profile (in blue). **e)** Macroscopic YM values reported in the literature[31,33,34] on halide perovskite single crystals of different compositions (each containing a single halide and a single monovalent cation). These values are comparable to the spatially averaged values obtained from our $Cs_{0.05}MA_{0.17}FA_{0.78}Pb(I_{0.83}Br_{0.17})_3$ perovskite thin films. Histograms of the YM values across the whole image shown in panel b) (in black) and separately for the regions identified as bulk (blue) or as mechanical boundaries (red) ($p<1e^{-4}$, unpaired two-tailed t-test). **f)** Compositional map showing normalized Br:Pb ratio extracted from nXRF peak intensities. The colour scale is linear, with darker areas representing regions with lower amount of Br. All scale bars correspond to 500 nm.

Using the topography images in Figure 1a, we identify large domains (associated with height local maxima) that are defined as morphological grains separated by grain boundaries (associated with height local minima). The lateral size of these morphological grains ranges between one and several hundreds of nanometers, which is similar in size to the morphological grains measured by scanning electron microscopy (Figure S1). Interestingly, the nanomechanical map (Figure 1b) reveals substantial YM variations across the sample, ranging between 2 GPa and 25 GPa, and provides the first demonstration of local variations in the stiffness of perovskite films. Moreover, our measurements not only show that interfaces between morphological grains are generally higher in YM (inter-grain features) but also unveil smaller mechanical subdomains (intra-grain features). This observation is further highlighted in Figure 1c, where topography and YM maps are overlayed, and in Figure 1d, where the variation in height and YM in the sample across a line profile (as defined by the blue dotted line in Figures 1a,b,c) is displayed. Mechanical domains are enclosed by regions of higher YM that are denominated as mechanical boundaries hereafter. This behaviour is reproduced in multiple samples (see Figure S8), implying that this is a general property in these mixed-halide perovskite films. The existence of mechanical domains within morphological grains can potentially be linked with the recent observation of structural features within morphological grains, which can impact optoelectronic performance of these materials[21,45].

To quantify the variations in YM, we break the data obtained by QNM-AFM down to values acquired from bulk areas and those acquired from the mechanical boundaries, based on the YM maps. We find spatially averaged values of YM to be 9.1 ± 2.6 GPa but observe significant differences between the YM in bulk sites (6.9 ± 1.3 GPa) and the YM at the mechanical boundaries (12.5 ± 2.9 GPa) (Figure 1e). In order to identify the origin of such mechanical variations, we explore how they are connected to compositional heterogeneities that arise during the thin film fabrication[25,46]. Although there is limited literature on the mechanical properties of halide perovskites, it has been reported that the macroscopic YM of these materials strongly depend on the halide content (summarized in the top panel in Figure 1e)[31,33,34]. Specifically, the pure Br-based perovskites show significantly larger YM values than their I-based counterparts: 17.5 – 19.6 GPa for $MAPbBr_3$ compared to 10 – 14.3 GPa for $MAPbI_3$[33],[47,48]. This difference in YM has been proposed to be associated with the higher Br-Pb bond strength in comparison to that of the I-Pb bond[35]. Moreover, Hutter *et al.* have reported macroscopic YM values for the I-Br mixtures that lie between those of the pure Br and I samples[47]. Although these macroscopic YMs were obtained by invasive techniques such as indentation, they correlate well with the spatially averaged values measured across our mixed-halide films. We hypothesize that local halide content has a direct influence on the local YM of mixed-halide films[47]. Indeed, Figure 1e shows that the bulk YM values are very close to the reported values for the pure I-based perovskites, while the values at the mechanical boundaries tend to be comparable to those for pure Br-based perovskites. To validate this proposition, the Br-to-Pb content ratio in our films is mapped using synchrotron-based scanning nXRF microscopy with a spatial resolution of ~50 nm. Figure 1f shows a distinct Br content heterogeneity in the films, with Br accumulation in specific sites that appear to be grain interfaces[46] and some bulk grains, following a similar trend as that of the mechanical properties. We therefore conclude that many of the mechanical boundaries (either morphological grain boundaries or sub-grain features) present higher Br content while most of the remaining material is consistent with having an I-rich $Cs_{0.05}MA_{0.17}FA_{0.78}Pb(I_{0.83}Br_{0.17})_3$ formulation, as shown in the YM maps.

It is known that light soaking induces halide migration in perovskite thin films, an effect that has a large influence on the performance of perovskite-based optoelectronic devices[26]. Given the connection between the local mechanical properties and local halide content in mixed-halide perovskites just revealed, these properties are expected to evolve when exposing the films to light. We perform correlative QNM-AFM and FLIM measurements on $Cs_{0.05}MA_{0.17}FA_{0.78}Pb(I_{0.83}Br_{0.17})_3$ films after light soaking them for 2 and 12 hours under continuous white-light illumination with an intensity of ~3750 mW/cm$^2$ (**Figure 2**). This mild illumination condition, approximately 3.5 times the intensity of natural sunlight on the Earth's surface, allows us to evaluate the impact of light on the mechanical and photophysical properties of the films without damaging the material. QNM-AFM maps acquired before (Figure 2a) and after (Figure 2b) illumination show that the topography of the film exhibits small changes in height after two hours of continuous illumination in laboratory ambient conditions. However, the corresponding YM maps reveal much larger modifications in the local mechanical properties of the sample (Figure 2c,d), with an overall reduction in the number of mechanical boundaries. This effect leads to a merging of adjacent mechanical domains and an increase in their effective size, resulting in a mechanical homogenization over the film after light soaking. Figure 2e displays a significant decrease in the spatially averaged YM from 10.3 ± 2.7 GPa to 7.6 ± 1.4 GPa after 2 hours of light soaking, with no further significant changes noticeable after 12 hours of illumination (7.6 ± 1.5 GPa). The spatially averaged YM after light soaking is very close to the average calculated for the bulk (6.9 ± 1.3 GPa) in Figure 1e, indicating further that there is homogenization of the YM of the sample towards the lower values.

To provide another perspective on the evolution of the physical properties of the film, we performed correlative time-resolved PL imaging and AFM on the same scan region (Figure 2f and 2g, respectively). These figures show a larger area (5×5 µm$^2$) encompassing the (2×2 µm$^2$) areas in Figure 2a-d, where the QNM analysis was performed. Upon light soaking, these FLIM maps of the perovskite film show an increase in both luminescence intensity (photobrightening) and PL lifetime (Figure 2g). The PL lifetime changes originate from the extended white light soaking and not from the pulsed laser illumination to acquire maps (Figure S9). The changes in PL lifetime are also visualized using phasor plots[49,50], which provide a global graphical representation of time-resolved data in every pixel without needing to invoke any physical models (see Methods). Importantly, the phasor plot reveals a trajectory of all pixels towards longer PL lifetimes (successive movement to the left) upon 2 hours and 12 hours of sample illumination. This enhancement in the PL of the films in ambient atmosphere is in good agreement with the literature[26] under these illumination conditions, which is linked to photo-chemical reactions involving oxygen species, moisture and mobile halides/defects that passivate trap states. Here, these observations correlate well with a homogenization of the mechanical properties of the material.

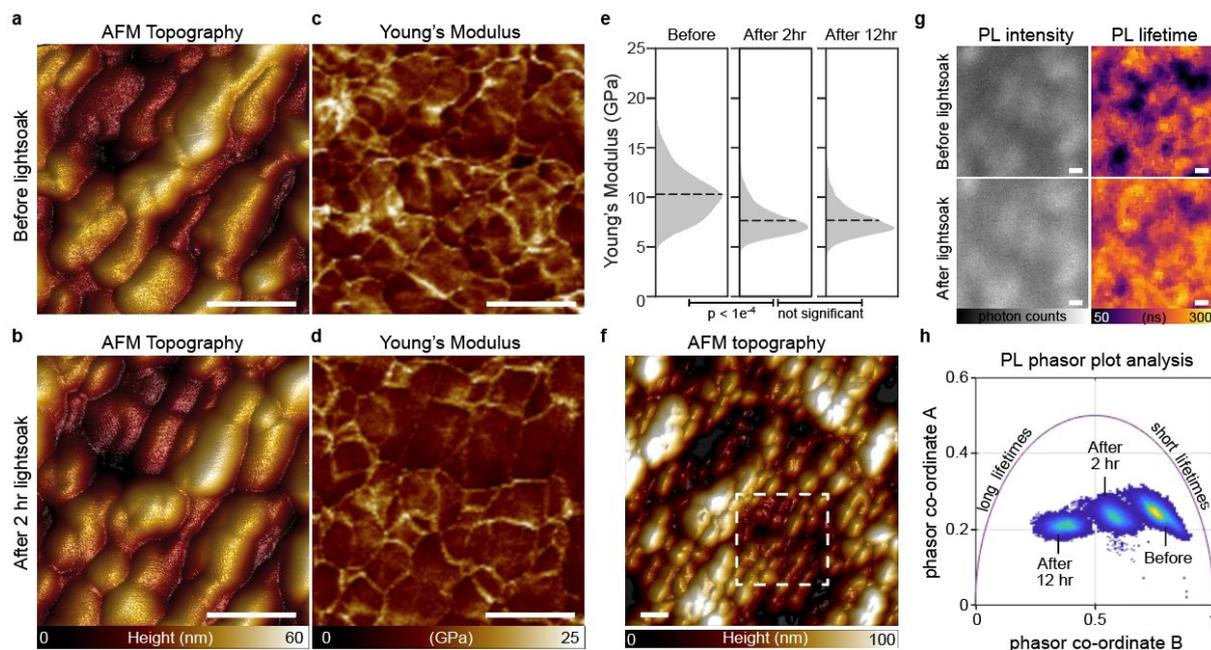

**Figure 2.** Atomic force microscopy and photoluminescence lifetime microscopy reveal changes in the mechanical properties and the emission lifetime of $Cs_{0.05}MA_{0.17}FA_{0.78}Pb(I_{0.83}Br_{0.17})_3$ perovskite thin films. **(a,c)** AFM topography and YM maps of the sample before light-soaking. **(b,d)** AFM topography and YM maps of the same area after 2 hours of white light soaking (~3750 mW/cm$^2$). While the topography of the sample remains largely unchanged, the YM values show significant homogenization. **(e)** Histograms showing a significant lowering and homogenization of YM values upon 2 hours and 12 hours of light soaking. **(f)** AFM topography map of a 5×5 μm$^2$ area of the perovskite film. Topography and YM maps of the 2×2 μm$^2$ region marked by dotted lines were investigated in (a-d). **(g)** Time-resolved PL maps of the same region as the AFM topography map in (f) shows photobrightening and an increase in PL lifetimes upon 2 hours of light soaking. **(h)** Phasor plot of time-resolved PL data showing a significant increase in PL lifetimes upon 2 hours of light soaking and a further increase upon 12 hours of light soaking for all areas in the illuminated region. All scale bars: 500 nm.

A pixel-by-pixel comparison and histograms of the PL lifetimes before and after illumination (**Figure 3**a) reveal a significant overall increase in the PL lifetime from 154 ± 44 ns to 197 ± 33 ns after light soaking ($p<1e^{-4}$, paired two-tailed t-test). The same analysis for YM values (Figure 3b) shows that there is a light-induced reduction in the YM of the sample, with the most pronounced reduction in the regions that are initially of high stiffness (highlighted by a dashed circle). We then separately quantify the YM associated with bulk areas before and after 2 hours of light exposure (Figure 3c). This analysis shows that the bulk areas that initially present low YM remain largely unchanged in terms of stiffness (7.8 ± 0.8 GPa and 8.0 ± 0.8 GPa before and after light soaking, respectively). However, the mechanical boundaries undergo a significant reduction in YM (13.8 ± 2.1 GPa before and 7.1 ± 0.8 GPa after light soaking), which can also been extracted from Figure 3b (see dashed circle). The YM of the mechanical boundaries tends to converge towards the spatially averaged YM of the bulk sites, effectively leading to a homogenization of the YM across the thin film. These results are seen in multiple batches, with statistical results from all samples ($n$ = 3 batches, 3 samples scanned in each) included in Figure 3c.

Taking the collective findings into account, we propose that our as-formed material contains Br rich regions[51]. This Br accumulation resides preferentially at the boundaries of morphological grains but also occurs in specific intra-grain regions (see top panel in Figure 3d). These Br clusters exhibit higher YM than the parent bulk regions defining mechanical boundaries that enclose mechanical domains in general smaller in size to morphological grains[52,53]. When light is introduced, it provides the required energy to promote ion migration, which happens preferentially at regions with a high fraction of Br (see bottom panel in Figure 3d). The mechanical properties of the film are homogenized due to the selective migration of previously clustered Br ions to become better intermixed with the bulk mixed-halide regions. Concomitantly, oxygen and light induced passivation processes involving ion migration reduce the deep trap density and non-radiative recombination processes in the films. This collective effect leads to improved optoelectronic properties expressed as an overall enhancement of the PL lifetimes measured in the films [25,54],[26,55].

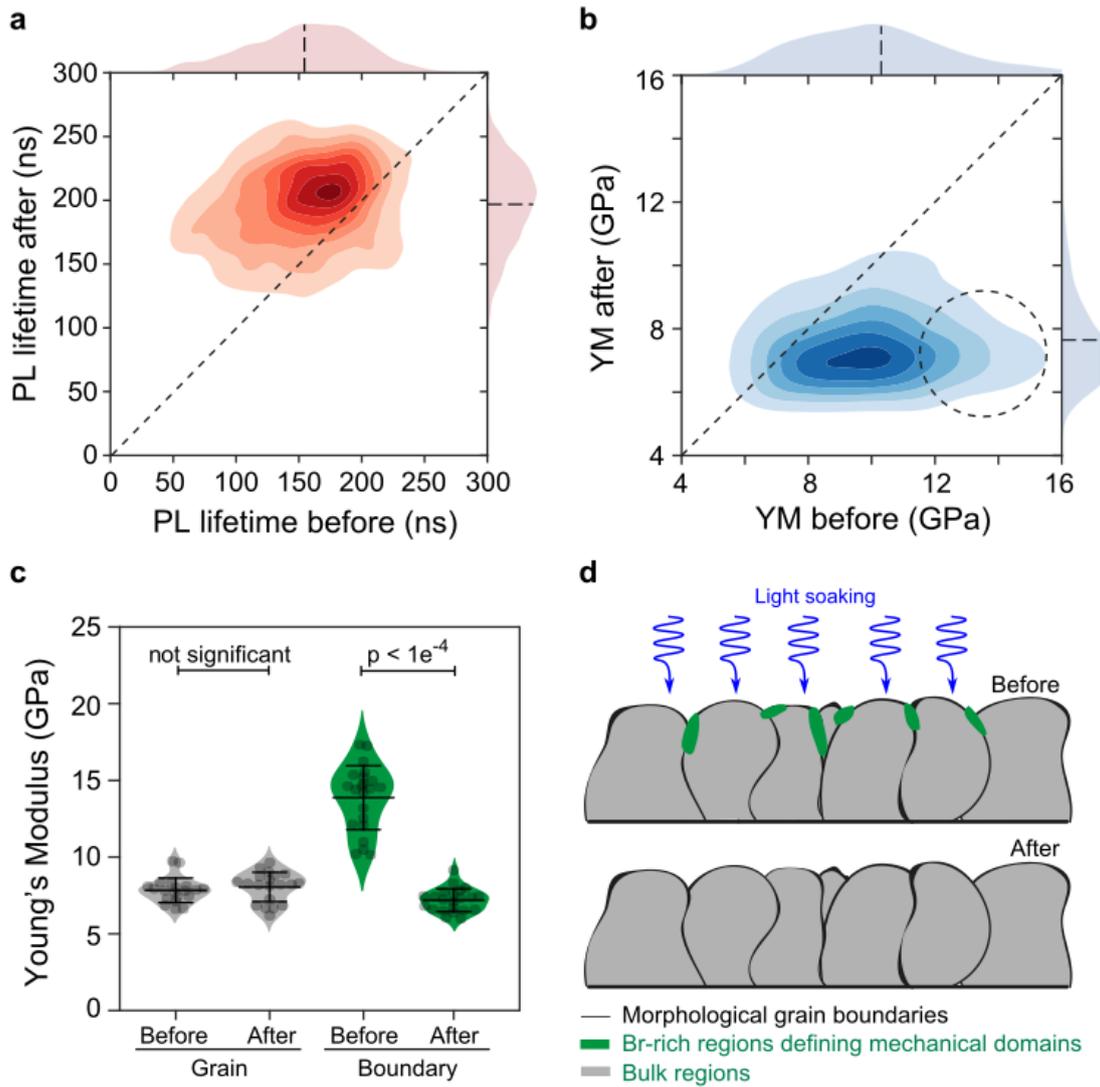

**Figure 3.** Material and photophysical properties of Cs$_{0.05}$MA$_{0.17}$FA$_{0.78}$Pb(I$_{0.83}$Br$_{0.17}$)$_3$ perovskite thin films before and after light soaking. **(a)** Pixel-by-pixel comparison and histograms of the PL lifetimes show a significant overall increase in the PL lifetimes after 2 hours of light soaking with white light at ~3750 mW/cm$^2$. **(b)** Pixel-by-pixel comparison and histograms of the YM show an overall reduction, especially of the higher stiffness values (dashed circle), after 2 hours of light soaking. **(c)** Distinct softening of the mechanical boundaries is observed after 2 hours of light soaking. The stiffness of the bulk regions of the morphological grain does not change. **(d)** Schematic representation of the cross-section of a triple cation mixed-halide perovskite film exhibiting several grains. Light-soaking induces a motion of the halides, leading to a homogenization of the halide (I$^-$/Br$^-$) ratio. In particular, it leads to diffusion of smooths out the large Br-rich clusters into the bulk, thus inducing a decrease of the YM values measured at the mechanical boundaries.

In conclusion, our study reveals that the mechanical properties of triple cation mixed-halide films vary on the submicron scale. These heterogeneities in the YM arise from the presence of stiffer, Br-rich areas, forming mechanical boundaries that define mechanical domains of a smaller size than morphological grains. These local mechanical properties evolve upon light soaking, with the stiffer mechanical boundaries homogenizing toward the YM values of the bulk of the halide-intermixed perovskite areas. Photobrightening also occurs on a similar timescale as the film stiffness homogenization, suggesting that the mechanical and photophysical properties in mixed-halide perovskites have a common shared origin in compositional heterogeneities. This study adds to the current understanding of the physical processes governing the working principles of these materials at the nanoscale, demonstrating that their mechanical properties are variable and evolve under relevant *in operando* conditions.

*Experimental Section*

*Fabrication of perovskite thin films:*

Triple cation perovskite $Cs_{0.05}MA_{0.17}FA_{0.78}Pb(I_{0.83}Br_{0.17})_3$ samples were deposited on 200 μm thick coverslips. Precursor solutions were prepared by dissolving $PbI_2$ (1.1 M), formamidinium iodide (FAI, 1 M), methylammonium bromide (MABr, 0.2 M) and $PbBr_2$ (0.22 M) in anhydrous DMF:DMSO (4:1 volume ratio). CsI (1.5 M in DMSO) was then added to the precursor solution. We followed a two-step thin film deposition program at 1,000 and 6,000 rpm for 10 and 30 s, respectively, and added 100 μl of chlorobenzene 30 s after the start of the spinning routine. We annealed the films at 100 °C for 1 h. Perovskite synthesis and thin film deposition were carried out in a $N_2$-filled glovebox to prevent them from degradation. Samples were always transported in hermetically sealed chambers to prevent them from degradation before performing any morphological, structural, mechanical or photophysical characterization.

*Scanning Electron Microscopy:*

We employed an analytical field-emission scanning electron microscope (FEI Nova NanoSEM) to image the morphology of the perovskite films. We used an in-lens detector in secondary electron detection mode and set the electron beam energy at 5 kV.

*X-ray diffraction*

X-ray diffraction patterns of the films were obtained with a Bruker D8 advance equipment with a Copper focus X-ray tube (Ka: 1.54 Å). The scan range for 2θ was from 10º to 40º with a step size of 0.10º.

*Photoluminescence spectra*

Photoluminescence spectra of the films were acquired using a widefield microscope (IMA VISTM, Photon Etc.). A CW 405 nm laser was employed to excite the samples and emitted light was detected in in reflectance mode with a 20x objective. We employed a tunable Bragg filter to obtain the spectrally resolved maps from 650 nm to 850 nm with a step size of 2 nm.

*Atomic Force Microscopy:*

Topography and Young's Modulus maps were acquired using a Bruker Bioscope Resolve in PeakForce QNM mode and RTESPA 525 probes (Bruker AFM probes) with a nominal spring constant of 200 N/m and resonant frequency of 525 kHz. The spring constant and deflection sensitivity of the probes were calibrated using a sapphire standard sample prior to each measuring session and the tip radius was estimated to be 20-30 nm, using a Titanium sample of known roughness. Topography maps of 5x5um$^2$ areas were first acquired and 2×2 μm$^2$ areas within these larger scans were probed with nanomechanical mapping. 2×2 μm$^2$ areas were scanned with 256 lines each, and 256 force curves were acquired within each line (see Figure S5 for a representative force curve). The extension part of the force curves was fitted to a linearised Hertz model for each pixel using Nanoscope 9.1 (Bruker), and the Young's Modulus was calculated as described in SI Section *Extraction of Young's Modulus/ Curve Fitting*. We defined mechanical boundaries as clusters of 5 or more force curves with YM above the global average.

*Time-resolved photoluminescence and correlative AFM-PL imaging:*

Time-resolved photoluminescence spectroscopy measurements were performed on a home-built, confocal microscope setup equipped with a pulsed supercontinuum source with 2 MHz repetition rate (Fianium Whitelase SC-400-4, 6ps pulse widths), a 100x objective lens (LMPlanFL N, 100X air, 0.8 NA, Olympus, Germany) and a time-correlated single photon counting module (SPC-830, Becker&Hickl GmBH, Germany). A bandpass filter FF01-474/27 was used to select a narrow excitation band and a longpass filter BLP01-635LP (both from Semrock Inc., New York, USA) was used to filter PL emission from the sample. Photons were acquired for fifteen minutes to collect enough photons per pixel in each 256×256 time-resolved PL image. The excitation power was maintained below 2.5 µJ cm$^{-2}$ pulse$^{-1}$. According to previous references[10,56], this fluence is high enough to saturate the traps and locally create Auger non radiative effects. Nevertheless, the confocal nature of the measurement may mean that carriers quickly diffuse and therefore the concentration quickly dilutes down to trap limited regime (which likely reflects the lifetimes we measure). The time taken for the PL to decay to 1/e [10,56,57] of its maximum intensity was calculated as the apparent PL lifetime in each image pixel using a custom MATLAB script. This calculation avoids biases that stem from fitting time-resolved data to mono and biexponential decays. A model-free phasor plot analysis was also performed using SPCImage v7.4 (Becker&Hickl GmBH, Germany). In this analysis, the PL decay in each image pixel undergoes a Fourier transform at the laser angular frequency to yield a complex variable called the 'phasor' [49,50]. The phasor coordinates representing the imaginary part and the real part are denoted by A and B, respectively. Thus, each image pixel is plotted as a point in the 2D phasor plot. Monoexponential decays lie on the universal circle (blue line, Figure 2h) while complex PL decays as seen in our samples (which depend on trapping, bimolecular recombination and diffusion[58]) lie within the universal circle. Any changes to the PL lifetime in a group of pixels lead to a displacement of points across the phasor plot.

For correlative AFM-PL imaging, AFM scans were performed first. PL images were captured after flipping the sample upside down for imaging in the inverted microscope frame and scanning the excitation laser over the same perovskite surface (see Figure S4). The regions of interest from both measurements were registered using signals from markers (gold nanoparticles or scratches) that were easily distinguishable in both the AFM and PL images.

*Nano X-ray fluorescence (nXRF)*

nf experiments were performed at the Hard X-ray Nanoprobe at the I14 beamline at Diamond Light Source Ltd., Didcot, UK as described previously[59,60]. Briefly, horizontal mirrors shape an X-ray beam from an undulator source onto a secondary source. The beam then passes over a Si monochromator which selects the chosen beam energy of 20 keV. Finally, the beam is incident upon a pair of Kirkpatrick-Baez mirrors which focuses the beam down to a size of ~50×50 nm$^2$. Perovskite samples for nXRF characterization were deposited as decribed above in a PTAA covered[61] 200 μm thick, 7.5×7.5 mm$^2$ SiN grids with a 1×1 mm$^2$, 100 nm thickness window. The sample is placed at the focal point of the beam and is raster scanned across this point. X-ray fluorescence is then captured by a silicon drift detector (Rayspec).

The multidimensional nXRF data were analyzed using DAWN[62], and a combination of Python packages, including Hyperspy[63]. The intensity of the Br Kα line and the Pb Lα lines were normalized to their respective means across the entire sca*n* region. Then the normalized Br signal is divided by Pb to get the relative ratio of Br:Pb. We note that nXRF measurements do not induce any of the observed Br heterogeneities across the sample.[52]

*Light soaking:*

Light soaking was performed either for 2 hours or 12 hours using a white light LED source (Xlamp XP-E2 LEDs, Cree Inc.). The source provides 0.375 W power distributed in an area of 1cm x 1cm$^2$, leading to a total fluence of 3750 mW cm$^{-2}$. This is an illumination equivalent to

~3.5 suns (assuming solar illumination is 1050 mW cm$^{-2}$). To quantify the YM across the whole film before light exposure and after 2 and 12 hours of light exposure, 10,000 force curves were fitted for each condition. The LED used for light soaking was turned off during PL acquisition.

**Supporting information**
Supporting information is available from the Wiley Online Library or from the author.

*Acknowledgements*
This project has received funding from the European Union's H2020-MSCA-ITN-2016 research and innovation programme under the Marie Sklodowska-Curie Grant Agreement No. 722380 (SUPUVIR). This project has received funding from the European Research Council (ERC) under the European Union's Horizon 2020 research and innovation programme (grant agreement number 756962). MA acknowledges funding from the European Union's Horizon 2020 research and innovation programme under the Marie Skłodowska-Curie grant agreement No.841386. C.F.K acknowledges funding from the Engineering and Physical Sciences Research Council, EPSRC (EP/ H018301/1, EP/L015889/1); Wellcome Trust (089703/Z/ 09/Z, 3-3249/Z/16/Z); Medical Research Council MRC (MR/K015850/1, MR/ K02292X/1); MedImmune; and Infinitus (China), Ltd. SDS acknowledges support from the Royal Society and Tata Group (UF150033). GD acknowledges the Royal Society for funding through a Newton International Fellowship and the EPSRC (EP/R023980/1). K.F. acknowledges a George and Lilian Schiff Studentship, Winton Studentship, the Engineering and Physical Sciences Research Council (EPSRC) studentship, Cambridge Trust Scholarship, and Robert Gardiner Scholarship. S.M acknowledges an EPSRC studentship. T.A.S.D acknowledges a National University of Ireland Travelling Studentship. The authors acknowledge the Diamond Light Source (Didcot, UK) for providing beamtime at the Hard X-ray Nanoprobe at I14 beamline (proposal: sp20420) and Julia Parker and Paul Quinn for beamline assistance.

**Conflict of Interest**

SDS is a co-founder of Swift Solar Inc.

# Supporting Information

**Revealing nanomechanical domains and their transient behavior in mixed-halide perovskite films**

Ioanna Mela[a#], Chetan Poudel[a#], Miguel Anaya[b]*, Géraud Delport[b], Kyle Frohna[b], Stuart Macpherson[b], Tiarnan A. S. Doherty[b], Samuel D. Stranks[a,b]*, Clemens Kaminski[a]*

Dr I. Mela, C. Poudel, Dr. S. D. Stranks, Prof. C. F. Kaminski
Department of Chemical Engineering & Biotechnology, University of Cambridge, Philippa Fawcett Drive, Cambridge CB3 0AS, UK.
cfk23@cam.ac.uk

Dr. M. Anaya, Dr. G. Delport, Kyle Frohna, S. Macpherson, T. A. S. Doherty, Dr. S. D. Stranks
Cavendish Laboratory, University of Cambridge, JJ Thomson Avenue, Cambridge CB3 0HE, UK.
ma811@cam.ac.uk
sds65@cam.ac.uk

# These authors contributed equally.

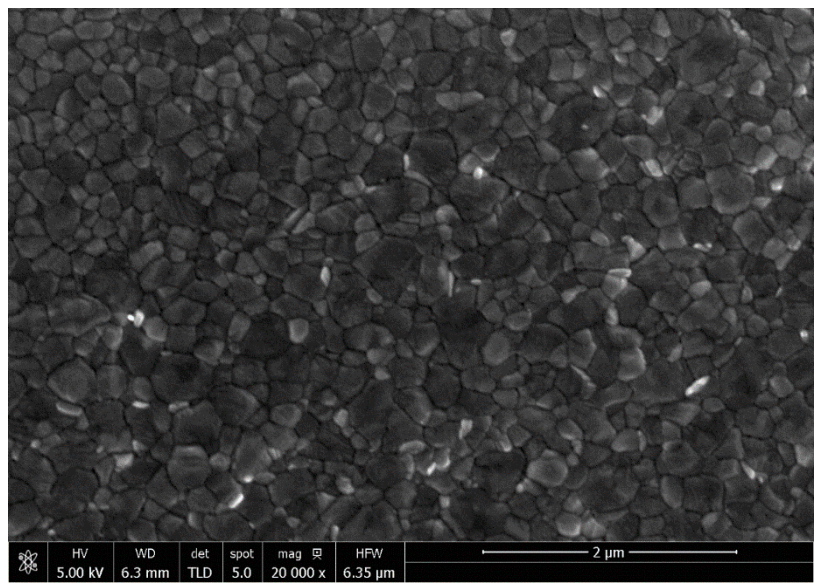

*Figure S1*: Scanning electron microscopy top view image of a $Cs_{0.05}MA_{0.17}FA_{0.78}Pb(I_{0.83}Br_{0.17})_3$ thin film. Scale bar: 2 μm.

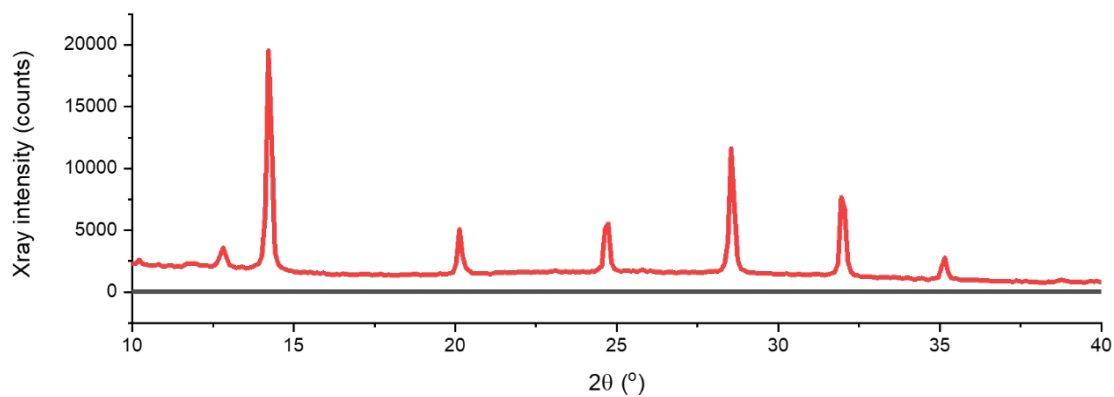

*Figure S2*: X-ray diffractogram of a $Cs_{0.05}MA_{0.17}FA_{0.78}Pb(I_{0.83}Br_{0.17})_3$ thin film.

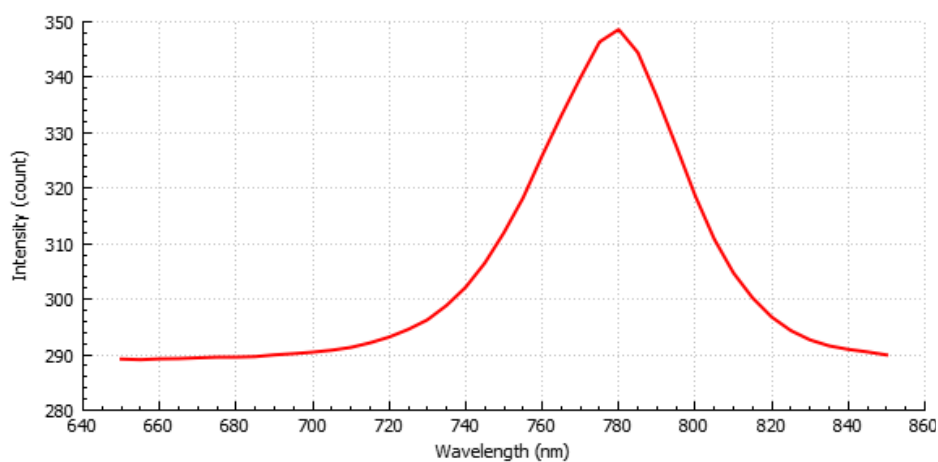

*Figure S3*: Photoluminescence spectrum of a $Cs_{0.05}MA_{0.17}FA_{0.78}Pb(I_{0.83}Br_{0.17})_3$ thin film.

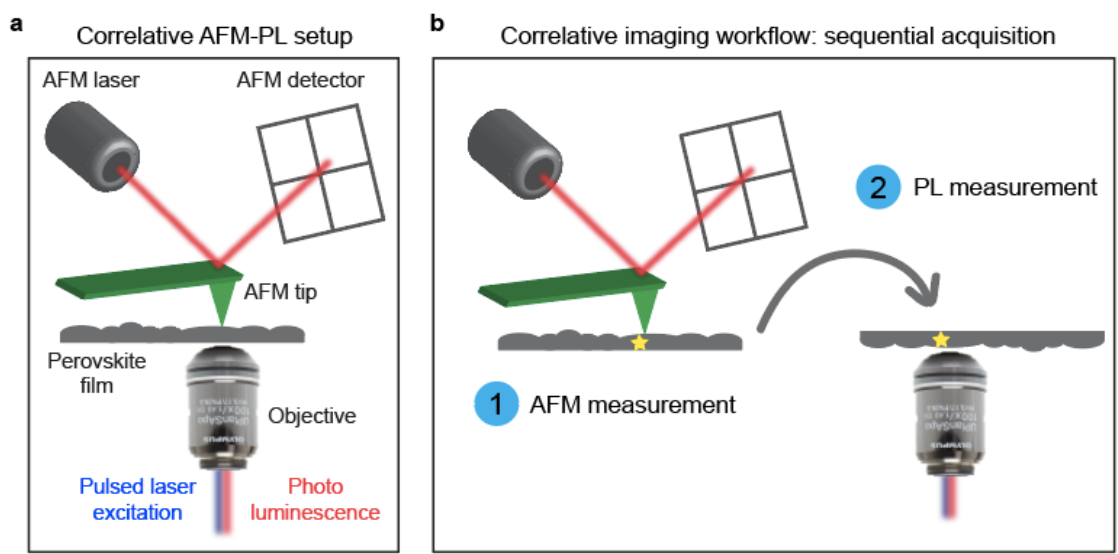

*Figure S4:* (a) Characterization setup based on a confocal microscope, in which correlative images of QNM-AFM and time-resolved PL can be obtained for the same region of interest. (b) Samples were flipped upside-down between QNM-AFM and PL scans as described in the Experimental Section.

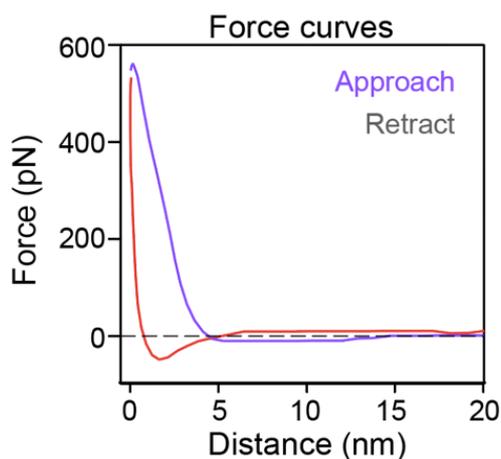

*Figure S5:* Typical force curves obtained for each pixel in the QNM-AFM images. Fitting these force curves to a Hertz model allows for extracting YM values per pixel.

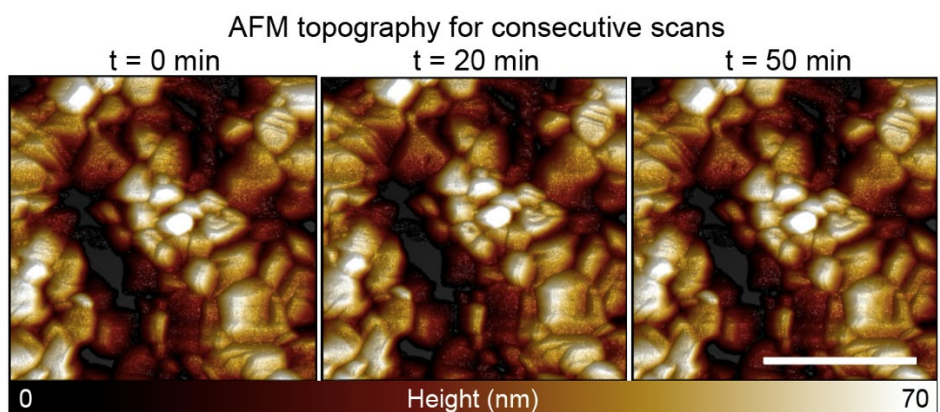

*Figure S6:* Topography of the perovskite films does not change over three consecutive QNM-AFM scans of the same field-of-view taken within an hour. Scale bar: 1 μm.

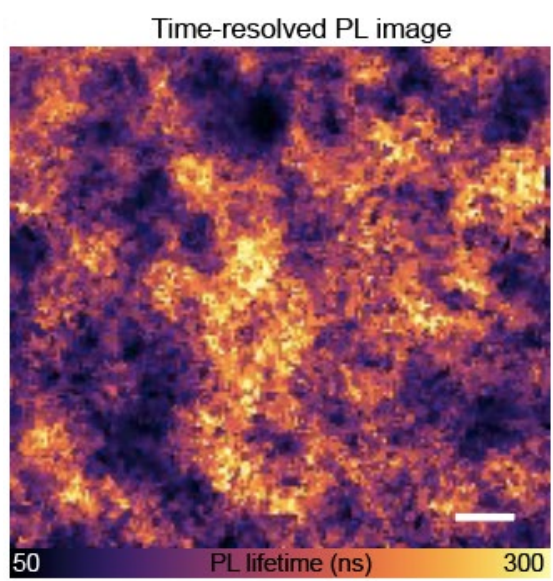

*Figure S7*: Typical time-resolved PL image of the triple cation perovskite thin film. The PL lifetime was calculated as the time taken for the PL to decay to 1/e of its maximum intensity. Scale bar: 1 μm.

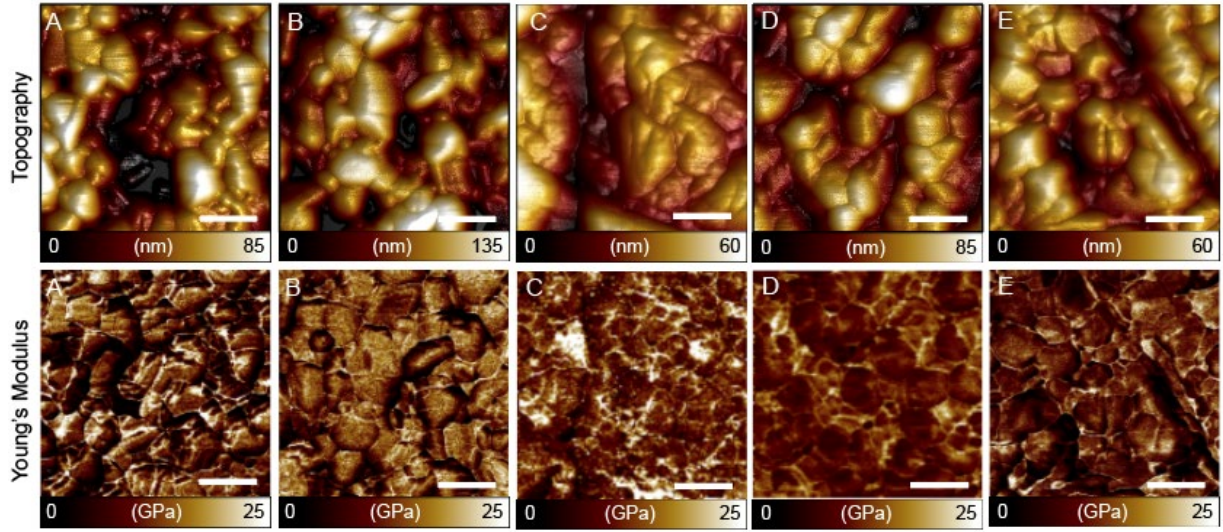

*Figure S8:* Topography and Young's Modulus maps obtained from several samples of $Cs_{0.05}MA_{0.17}FA_{0.78}Pb(I_{0.83}Br_{0.17})_3$ thin films. Scale bar: 500 nm.

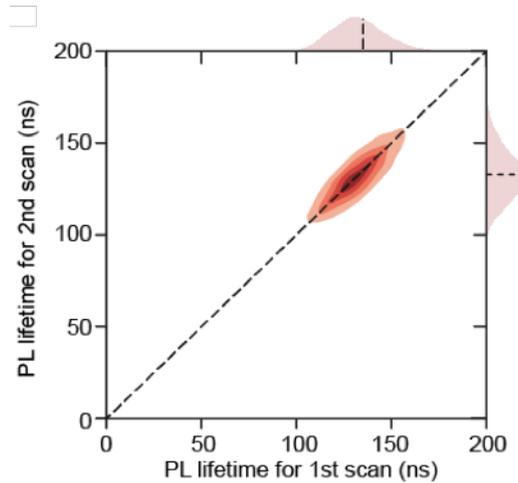

*Figure S9:* Pixel-to-pixel correlation and histograms show that the PL lifetime of the perovskite films does not change due to illumination from the pulsed laser over consecutive image scans of the same field-of-view taken within half an hour. The LED source used for light soaking was not in use in this case.

**Extraction of Young's Modulus/ Curve Fitting**

To calculate Young's Modulus (*E*), the first 5nm of the converted force vs. indentation curve were processed using the Hertz model. The Hertz model approximates the sample as an isotropic and linear elastic solid occupying an infinitely extending half space. It is also assumed that the indenter is non-deformable and that there are no additional interactions between indenter and sample. Besides, the Herz model assumes that the indentation is negligible in comparison to the sample thickness, i.e. the model is valid for small indentations, up to 5-10% of the total thickness of the sample. If these conditions are met, the Young's Modulus (E) of the sample can be fitted and calculated using the Hertzian model. In our case, we are using an extension to the Herz model, that describes a parabolic indenter geometry, and therefore the model is modified as follows:

$$F = \frac{4\sqrt{R_c}}{3} \frac{E}{1-v^2} \delta^{3/2}$$

Where $R_c$ is the radius of tip curvature, $v$ is the sample's Poisson's ratio, E is the Young's Modulus, and $\delta$ is the indentation depth.

The data obtained by atomic force microscopy indentation measurements are in the form of force against piezo displacement. Therefore, for the Herz model to be applied, the force curves need to be converted to force against tip separation. $\delta$ is calculated by subtracting the cantilever deflection from the distance that the cantilever is moved towards the sample. In this way, a force (load) —indentation curve is created. This curve consists of two parts, the loading part in which the tip is moving towards the sample and the unloading part in which the motion of the tip is the opposite (Figure S5). The approach (loading) curve has been used to extract the Young's Modulus in each case since it is free from any other potential interactions such as adhesion, that may influence the accuracy of defining the point at which the sample and tip come into contact.

It should be noted that for hard samples, such as the ones used in this study, the Oliver & Pharr model[64] has been used previously for data processing. However, the Oliver & Pharr model is better suited in cases where very large forces are applied to deform hard samples and result in elastic-plastic contact rather than the purely elastic contacts observed in our case. A recent study supports that the Hertzian and Oliver & Pharr analysis provide identical results for purely elastic samples and parabolic indenters[65]. An interesting field for future work would be the adaptation of the existing models to increase the precision of nanomechanical mapping analysis of thin perovskite films.

RTESPA – 525 AFM probes (Bruker AFM probes) were used. Typical length, width, and thickness of these cantilevers are 125, 40, and 5.75 μm, respectively, nominal force constant k = 200 N/m, and nominal resonant frequency in air 525 kHz. The nominal tip radius is reported as 8 nm; however, when calibrating the tip radius with the use of reference samples (indirect method, on HOPG), the rip radii were found to vary between 20 – 35 nm. The tip shows a rotated geometry.